# A Native Hawaiian-led summary of the current impact of constructing the Thirty Meter Telescope on Maunakea

**Sara Kahanamoku\***, *Department of Integrative Biology & Museum of Paleontology, University of California, Berkeley, Berkeley, CA 94720*
**Rosie 'Anolani Alegado PhD**, *Department of Oceanography, University of Hawai'i Mānoa, HI 96822*
**Aurora Kagawa-Viviani**, *Department of Geography and Environment, University of Hawai'i at Mānoa, HI 96822*
**Katie Leimomi Kamelamela PhD**, *Akaka Foundation for Tropical Forests, Kamuela, HI 96743*
**Brittany Kamai PhD**, *California Institute of Technology, Pasadena, CA 91125*
**Lucianne M. Walkowicz PhD**, *Astronomy Department, The Adler Planetarium, Chicago, IL 60605*
**Chanda Prescod-Weinstein PhD**, *Department of Physics & Astronomy and Department of Women's & Gender Studies, University of New Hampshire, Durham, NH 03824*
**Mithi Alexa de los Reyes**, *Department of Astronomy, California Institute of Technology, Pasadena, CA 91125*
**Hilding Neilson PhD**, *Department of Astronomy & Astrophysics, University of Toronto, Toronto, ON, M5S 3H4*

*\*Corresponding author; sara.kahanamoku@berkeley.edu*

**Disclaimer: Addressing every recommendation in this white paper <u>does not constitute any form of consent from Native Hawaiians for TMT or for a Maunakea lease permit</u>. Our recommendations are minimum first steps that can be undertaken to begin a process of building an iterative and equitable relationship with Native Hawaiians.**

**Executive summary**

Maunakea, the proposed site of the Thirty Meter Telescope (TMT), is a lightning-rod topic for Native Hawaiians, Hawai'i residents, and the international astronomy community. In this paper we—Kanaka 'Ōiwi (Native Hawaiian) natural scientists and allies—identify historical decisions that impact current circumstances on Maunakea and provide approaches to acknowledging their presence. Throughout this paper, we expand dialogue and inform actions utilizing a native Hawaiian concept known as kapu aloha, which "helps us internationalize our thoughts, words and deeds without harm to others"[1,2]. **Our aim is to provide an Indigenous viewpoint centered in Native Hawaiian perspectives on the impacts of the TMT project on the Hawaiian community.**

In this paper we provide a summary of the current Maunakea context from the perspective of the authors who are trained in the natural sciences (inclusive of and beyond astronomy and physics), the majority of whom are Native Hawaiian or Indigenous. We highlight three major themes in the conflict surrounding TMT: 1) physical demonstrations and the use of law enforcement against the protectors of Maunakea, nā kia'i o Mauna-a-Wākea; 2) an assessment of the benefit of Maunakea astronomy to Native Hawaiians; and 3) the disconnect between astronomers and Native Hawaiians. We close with general short- and long-term recommendations for the astronomy community, which represent steps that can be taken to re-establish trust and engage in meaningful reciprocity and collaboration with Native Hawaiians and other Indigenous communities. Our recommendations are based on established best principles of free, prior, and informed consent and researcher-community interactions that extend beyond transactional exchanges. **We emphasize that development of large-scale astronomical instrumentation must be predicated on consensus from the local Indigenous community about whether development is allowed on their homelands**. Proactive steps must be taken to center Indigenous voices in the *earliest* stages of project design.

To this end, we provide seven major recommendations for ongoing and future astronomy research on Maunakea and other sacred Indigenous lands:
1. **Immediately halt Thirty Meter Telescope progress and work with Native Hawaiian cultural knowledge holders to restart dialogue with the goal of obtaining informed consent.** Construction cannot proceed without consent from Native Hawaiians; the astronomical community must be willing to accept that a "no deal" outcome may ultimately be requested by Native Hawaiians or the State of Hawai'i.
2. Establish a Cultural Impact Assessment process that is viewed as legitimate by standards determined within the Native Hawaiian community.
3. Require that every observational astronomer learn Hawaiian history and culture, regardless of whether they are physically present in Hawai'i.
4. Establish equitable, iterative dialogue with Native Hawaiians.
5. Invest in support for Native Hawaiian astronomy students.
6. Develop astronomy-specific ethical guidelines and accountability structures.
7. Funding agencies must hold PIs accountable for the research environments they create.

## 1. Background

Maunakea is Kanaka ʻŌiwi ancestral land. The Mauna—also known as Mauna Kea and Mauna-a-Wākea[i]—is one of the most sacred places in the Hawaiian Islands, and stands as a place of worship, an ancestor to Native Hawaiians, and a piko (umbilicus, or site of convergence) for the lāhui Hawaiʻi (Hawaiian nation)[3,4]. Maunakea is a central element in the Kumulipo, a cosmological chant structured around the observation of environmental and celestial patterns[3]. The Mauna's position as an elder sibling to the Hawaiian people in the Kumulipo illustrates a central concept in Hawaiian culture: aloha ʻāina, or a familial love for and commitment to sustaining the land, drives the foundational duty to value land. In perpetuating aloha ʻāina, Native Hawaiian well-being and the well-being of the land are interdependent; neither can exist without the other.

Kū Kiaʻi Mauna is a Native Hawaiian hui (collective) whose goal is to protect Maunakea by preventing construction of the Thirty Meter Telescope (TMT) on the summit, and center a diversity of Native Hawaiian voices within all activities in Hawaiʻi. As this paper is submitted, kiaʻi (protectors[ii], both Native Hawaiian and non-Hawaiian) remain present near a kipuka, a volcanic cinder cone outcropping, named Puʻu Huluhulu. Their non-violent actions, led by respected Native Hawaiian kūpuna (elders), were met with aggressive opposition from state-level officials, who endeavored to displace people through arrests rather than engaging in respectful dialogue. These events have catalyzed state-wide and world-wide movements centered on a fundamental question: **do Indigenous people have the power to decide what happens to their own homelands?** Throughout the life of the TMT project, this question has reverberated deeply within the Hawaiian community.

Tensions between Native Hawaiians and astronomers arise from Maunakea's status as one of the best places in the world for ground-based astronomy. The pristine atmospheric conditions present on Maunakea has led to the construction of 13 telescope complexes, which produce the majority of data collected in the Northern Hemisphere. Yet astronomy's presence on Maunakea has directly resulted and benefited from the United Sates (U.S.) takeover of Hawaiʻi and appropriation of the personal lands of the last reigning monarch of the Hawaiian Kingdom (crown lands, or "ceded lands"). Current efforts to protect Maunakea have generated renewed attention around the United States' role in the illegal overthrow of Queen Liliʻuokalani in 1893, when the U.S. Minister and military representatives conspired with American and European businessmen to persuade armed U.S. forces to invade the sovereign Hawaiian Kingdom. These unlawful actions towards an independent nation established a provisional government that eventually transitioned into the State of Hawaiʻi in 1959, and led to the taking of Hawaiian lands, cultural resources, and self-determination with long-lasting detrimental impacts on Hawaiian political, social, economic, and value systems[5].

Dispossession of Native Hawaiians from their homelands remains a primary issue threatening Hawaiian identity and well-being, and the separation of Hawaiian cultural practitioners from spaces such as Maunakea heightens this intergenerational trauma. The relationship between institutional astronomy and Native Hawaiians has been unbalanced and prioritized research since the construction of the first telescopes in the late 1960s, and uneven dynamics on the Mauna are encapsulated in the viewpoints held by some members of these communities. Many astronomers who use data from telescopes on Maunakea view their work as inherently nonviolent and in the common interest of humanity. In contrast, many Native Hawaiians assert that their Indigenous rights to self-determination are under siege, while astronomers directly benefit from the disenfranchisement of Hawaiians[3,6]. **At the heart of this**

---

[i] These names hint at the cultural significance of Maunakea: Mauna-a-Wākea means "Mountain of Wākea," named for the sky deity in Hawaiian culture, whose daughter Hoʻohōkūkalani represents the stars that astronomers hope to view from the mountain's peak.
[ii] Frontline Native Hawaiians and allies are self-described as "protectors, not protestors" and kiaʻi mauna (guardians of the mountain). We use the term kiaʻi to respect this self-identification[4].

**disconnect is a question of power: the choices made on Maunakea reflect who is granted authority to make decisions for Native Hawaiians, contested lands, and for the nature and terms of cultural practice on these lands.**

The Native Hawaiian community is not of one mind as to whether TMT will truly benefit generations of Native Hawaiian people. Kiaʻi are unconvinced that the project stakeholders have demonstrated adequate accountability in upholding their responsibilities or promises around Maunakea management[7], and maintain that the project only perpetuates the desecration of sacred sites and significant environmental impacts[iii]. Supporters of Imua TMT, however, hope that moving the TMT forward will provide access to unique education and employment opportunities for residents of the State of Hawaiʻi. **The multiplicity of viewpoints demonstrates that there is no broad consent for TMT among Native Hawaiians**; a significant amount of work is still required to reach a resolution.

Native Hawaiians have the right, as expressed in the United Nations Declaration on the Rights of Indigenous Peoples (UNDRIP)[8], "to self-determination." Of particular importance in the UNDRIP framing of Indigenous rights is the **requirement that projects receive explicit, informed, and *ongoing* consent from Indigenous peoples**—we emphasize that this requires more than involving Indigenous individuals in consultation. The recent developments on Maunakea, as well as the history of legal challenges to TMT and earlier endeavors (*e.g.*, the Keck Outriggers[iv]), demonstrate that TMT currently lacks consent from the local Indigenous community. As such, **the TMT project must reconsider its position: is there a path forward, or should they withdraw and consider an alternate location?**

Though these questions are difficult, astronomers must consider their obligations to the Indigenous people of Hawaiʻi if they hope to do astronomy on Maunakea in an ethical and non-violent manner. Native Hawaiian cultural knowledge holders, including those not affiliated with the fields of astronomy, must consent—not merely be consultants to—further development. Inherent in the consent process is the ability to lead in decision-making.

Native Hawaiians have a right to decide for ourselves the future of our Mauna.

## 2. Impacts of the Thirty Meter Telescope project on Native Hawaiians

Controversy surrounding telescopes on Maunakea is not novel. Native Hawaiians have been vocal about the presence of structures on the Mauna since the inception of the Maunakea Observatories. Evidence of this exists in many different forms, including in litigation, op-eds and media interviews, attendance at local and state town halls, civil disobedience and arrests, social media exchanges, musical arrangements, the creation of new oli (chants) and moʻolelo (stories, legends), and renewed efforts for formal recognition and Hawaiian sovereignty. Though Native Hawaiians have diverse viewpoints and opinions—

---

[iii] *Re Conservation District Use Application for TMT, SCOT-17-0000777, Dissenting Opinion (Nov. 9, 2018):* "The Board of Land and Natural Resources (BLNR) grounds its analysis on the proposition that cultural and natural resources protected by the Constitution of the State of Hawaiʻi and its enabling laws lose legal protection where degradation of the resource is of sufficient severity as to constitute a substantial adverse impact. Because the area affected by the Thirty Meter Telescope Project... was previously subjected to a substantial adverse impact, the BLNR finds that the proposed TMT project could not have a substantial adverse impact on the existing natural resources. Under this analysis, the cumulative negative impacts from development of prior telescopes caused a substantial adverse impact; therefore, TMT could not be the cause of a substantial adverse impact. As stated by the BLNR, TMT could not "create a tipping point where impacts became significant." **Thus, addition of another telescope—TMT—could not be the cause of a substantial adverse impact on the existing resources because the tipping point of a substantial adverse impact had previously been reached**" (emphasis added).

[iv] For a timeline of Maunakea legal actions, see http://kahea.org/issues/sacred-summits/timeline-of-events.

like other social groups, we do not operate as a monolith—support for TMT has significantly decreased among Hawaiʻi's Indigenous people throughout the life of the project. In contrast to an oft-cited survey, which suggested that 72% of Native Hawaiians were in support of TMT (March 2018, N = 78)[v], more recent assessments with larger sample sizes show that closer to **27% of the Native Hawaiians polled were in support of TMT** (September 2019, N = 400)[vi]. Given this shift in opinion, a clearer assessment of Hawaiian community views is imperative.

## 2.1 2015 and 2019 demonstrations: State law enforcement used against kiaʻi

While Native Hawaiians have organized opposition to the TMT since the earliest phases of the project's planning and approval process[3], there have been two major physical conflicts to date. In April 2015, kiaʻi physically blockaded the road to prevent movement of construction equipment up the Mauna; 31 were arrested. The hundreds who took a stand over the subsequent demonstrations on the Mauna and at the flagship University of Hawaiʻi at Mānoa campus forced Governor Ige to temporarily postpone the project. In December 2015, the Hawaiʻi Supreme Court invalidated TMT's construction permits, ruling that the project failed to meet the requirements of due process.

Following a protracted legal battle, TMT permits were approved in July 2019[iii]. Governor Ige closed public access to Maunakea, and construction was to commence on July 15. Native Hawaiians again organized in protection of Maunakea by establishing a Puʻuhonua (place of refuge) at Puʻu Huluhulu to serve as the long-term center of demonstrations. Hundreds of Native Hawaiians and supporters gathered at the base of Maunakea, where eight kiaʻi chained themselves to a cattle grate--once again placing their bodies on the line to prevent the movement of construction equipment. On July 16, 2019, revered kūpuna (elders cultural knowledge holders) took to the frontlines and 33 were arrested. The sacrifices of kūpuna galvanized a Native Hawaiian movement grounded in kapu aloha (a reverence for love) and drew individuals by the thousands from around the world. In the month of August 2019 alone, 15,000 people visited the Puʻuhonua (a number comparable to ~8% of the population of Hawaiʻi Island, where Maunakea is located). Governor Ige declared a temporary state of emergency which granted heightened power to law enforcement agencies and **provided the possibility of deploying the National Guard in direct support of a research endeavor.**

As this paper is submitted, construction has not progressed, yet non-violent civil engagement has not slowed: over 10,000 people marched through Honolulu in October 2019, and hundreds of kiaʻi remain at Puʻuhonua o Puʻuhuluhulu, costing the State of Hawaiʻi $11 million on security. **The presence and threat of law enforcement by the State of Hawaiʻi against kiaʻi adds tension to ongoing efforts to affirm the sacredness of Maunakea, and deepens the harmful impacts of the TMT on Native Hawaiians.** To date, hundreds of astronomers have signed an open letter[vii] denouncing arrests; this letter "ask[s] that the community pause and consider what it means that, armed or not, the military and the police have become involved in the project's deliberations with the protectors of Maunakea."

## 2.2 Distinct worldviews

The controversy surrounding astronomy on Maunakea, including the TMT process, must be positioned within a historical context. Frustrated communications result in part from fundamental differences between the worldviews held by some astronomers ("mainstream science") and those held by Native Hawaiian cultural practitioners ("Indigenous knowledge systems"). Indigenous nations and peoples have explored the world and universe for far longer than western astronomers have had telescopes; while these knowledges are not uniform, Indigenous knowledge systems build upon a deep connection with the land, the water, and the sky through consistent observations. These systems utilize axioms that differ from traditional western science, and as a result may center different values than those of mainstream science.

---

[v] Summary of 2018 poll: https://www.scribd.com/document/374711098/The-Hawaii-Poll-March-2018-TMT
[vi] Summary of 2019 poll: https://www.scribd.com/document/427307684/Hawaii-Poll-TMT-Questions-Sept2019
[vii] Open letter opposing criminalization of Maunakea protectors:
https://docs.google.com/document/d/1YR8M4eboRjJSsfvVtmukb6dDgUonDBdmj9AU0h1rkmY

The public prioritization of the goals, values, and concerns of professional astronomers over those of the Indigenous inhabitants of Hawaiʻi insinuates that Native Hawaiian viewpoints on Maunakea are unimportant, or that only certain Native Hawaiian views are acceptable. As an example, portrayals of Native Hawaiians as "anti-science" have long been used in popular discourse regarding the movement for Maunakea: While some astronomers portray their science as "universally beneficial" to humanity[3], kiaʻi who stand in kapu aloha are portrayed as impediments to progress. To illustrate this, Native Hawaiian scholar Iokepa Casumbal-Salazar writes[3]:

> One scientist told me that astronomy is a "benign science" because it is based on observation, and that it is universally beneficial because it offers "basic human knowledge" that everyone should know… Such a statement underscores the cultural bias within conventional notions of what constitutes the "human" and "knowledge." In the absence of a critical self-reflection… the tacit claim to universal truth reproduces the cultural supremacy of Western science as self-evident. Here, the needs of astronomers for tall peaks in remote locations supplant the needs of Indigenous communities on whose ancestral territories these observatories are built… "Why would anyone oppose astronomy? Why are Hawaiians standing in the way of progress?" they ask. "Can't astronomers and Hawaiians coexist on the mountain?" These frames decontextualize the historical relations in which the TMT controversy has emerged and dehistoricize the struggle over land and resources in Hawaiʻi by vacating discourse on settler colonialism in favor of problematic claims to universality. When the opposition to the TMT is misrepresented as an arbitrary disregard for science, Hawaiians appear unreasonably obstinate.

The characterization of Hawaiians as, *e.g.*, "backwards" and "[unmoved] by logic" is discriminatory language, and should be interrogated as such when this language emerges from institutions of higher education. The 2015 demonstrations were described[8] as an "attack on TMT by hordes of Native Hawaiians who are lying about the impact of the project… and who are threatening the safety of TMT personnel." A tenured faculty member at the University of Hawaiʻi wrote[9] in 2015 that "in no way should we go back a few centuries to a stone age culture, with a few (illegitimate) Kahunas telling everyone how to behave." While these statements were denounced by pro-TMT groups, these types of comments continue to emerge from frustrated tenured physics and astronomy faculty[viii]. This sends the message that astronomers on Maunakea, and the astronomy community at large, are dismissive of raised concerns and see Native Hawaiians and supporters as subhuman. Further, because these raucous ideals typically come from senior faculty who are in positions of power and authority, they act to silence and alienate Indigenous astronomers, who are overwhelmingly in junior positions.

Paradoxically, while many astronomy departments and institutions undertake diversity initiatives (*e.g.*, establishing Codes of Conduct, focusing recruiting efforts at national conferences on underrepresented groups, *etc.*), the failure of astronomers to internalize Native Hawaiian assertions on the sacredness of Maunakea seriously renders these efforts hollow. Given that astronomers from historically marginalized groups may draw parallels between Native Hawaiian sovereignty struggles and their own commitments to their communities, suppression of nuanced opinions over TMT as well as dismissal of the struggles of Indigenous peoples are also an implicit dismissal of their concerns. Minoritized astronomers may view

---

[viii] Public archive of relevant UH emails: https://listserv.hawaii.edu/cgi-bin/wa?A1=ind1908&L=UHARI-L&X

abusive language directed at Native Hawaiians as echoing slurs (racial and otherwise) that have been directed at them.

## 2.3 Do Native Hawaiians significantly benefit from astronomy on Maunakea?

Within the last few years, a number of successful Hawaiian-centered and Hawaiian-led education programs have been piloted at the Maunakea Observatories[10] and ʻImiloa[11] in an attempt to push a "collaboration with integrity" model that combines Indigenous knowledge systems with mainstream science[12]. However, this outreach is targeted at a select few; the lack of outreach programs in marginalized rural communities only compounds the decades-long view that the University and state of Hawaiʻi and astronomers are unresponsive to community concerns regarding the development and management of the Maunakea summit. Programs such as A Hua He Inoa are more effective at bringing Hawaiian language to the global astronomy stage than as outreach and service in alignment with Native Hawaiian and local needs, unevenly distributing their benefits. The conflict over TMT construction on Maunakea highlights the pressing need for reciprocal dialogue with the Native Hawaiian community. Yet instead of engaging, the University leadership and astronomy community have stepped back to allow state and county law enforcement agencies to intervene on behalf of private astronomical interests. Moving forward, cultural programming and other community-based efforts should center on Hawaiian values, including aloha ʻāina, and be conducted with meaningful and iterative dialogue about the issues that are tearing at the social fabric of Hawaiʻi and beyond. A substantial amount of work is required to establish trust, develop content, and produce accountability, and significant funding must be allocated to meaningfully facilitate this work.

Both in the Hawaiian Islands and on a broader scale, astronomy education and public outreach relies on narratives that curiosity about space is a uniting "human" experience (*e.g.*, "...that is what makes astronomy beautiful. To study something—not because we're looking to gain anything in particular, but out of sheer curiosity—is what makes us human"[13]). However, these notions are antithetical to the colonial behaviors the astronomy community has engaged in and reinforced by denying the humanity of Native Hawaiians for the past 50 years. In this context, outreach efforts claiming a shared humanity are not only unconvincing, they ultimately undermine the perception of astronomers' integrity. Critically, astronomy funding is dependent on this perception: "the generous public support for NASA's astronomy research stems largely from astronomers' success in making the fruits of their research accessible and appealing to many people"[14]. Indeed, astronomers enjoy being able to share the results of their research and many now engage in education and public outreach as a central career path[15], evidence of broad support for these efforts within the astronomy community. However, astronomers wishing to share the results of their scientific efforts cannot expect to have receptive audiences indefinitely: millions of people across the world have witnessed our elders being arrested in July 2019[ix] on social media and major news outlets. **These visual records of astronomer complicity with state violence have indelibly marred astronomy's claims to a shared humanity.** If astronomers wish to return to sharing the results of their research with a receptive public, and to the opportunities for public support (both intangible and financial) they have previously enjoyed, they must find a path forward that centers the legitimate concerns of Native Hawaiians.

---

ix See, for example, this video on the arrests: https://twitter.com/karaokecomputer/status/1151821442247356417

## 3. Near-term recommendations: address long-standing issues in the relationship between astronomers and Native Hawaiians

Here we provide seven recommendations for improving the relationship between astronomers and Native Hawaiians over the next decade. Due to the urgency of the Maunakea situation, we focus these recommendations on the TMT case, but suggest that they can be applied toward improving astronomy's relationship with other Indigenous groups (given that they are adapted for each group's unique historical contexts and worldviews). We note that these recommendations are not novel; a large body of resources exist to facilitate research with and among Indigenous peoples, and should be leveraged in future interactions[16].

### 3.1. Immediately halt TMT progress and work with Native Hawaiian cultural knowledge holders to restart dialogue with the goal of obtaining informed consent. Be willing to accept a "no deal" outcome.

**Problem 1:** The TMT project does not need to remain a source of conflict between astronomers and Native Hawaiians. The looming threat of arrest is a form of violence towards kia'i, and if the project is not reassessed, other threats—to physical safety and well-being—may escalate as winter approaches. **Kia'i are extremely committed to the Mauna, and are willing to die to stop the construction of TMT.**

**Recommendation 1: Immediately halt progress on TMT (including construction attempts) until clear, informed, and ongoing consent is obtained from a diverse community of Native Hawaiians**, including the kūpuna cultural knowledge holders who are most familiar with the Mauna's historical and spiritual significance. Allow Native Hawaiians to negotiate a non-violent process to build consensus on appropriate next steps. Construction cannot proceed without broad consent from Native Hawaiians, and the astronomical community *must* respect that a "no deal" outcome may ultimately be requested by Native Hawaiians or State of Hawai'i.

### 3.2. Establish a Cultural Impact Assessment process that is viewed as legitimate by standards determined within the Native Hawaiian community.

**Problem 2:** The TMT project progressed through to its final stages in the face of vocal and repeated opposition from Native Hawaiian cultural knowledge holders, stakeholders, and community members. TMT strategic communications promote curated Native Hawaiian viewpoints not grounded in cultural practices connected to Maunakea though are masters in their own discipline and place.

**Recommendation 2:** Establish a Cultural Institutional Review Board (IRB) requirement for astronomy projects to formalize the requirement that all projects receive free, prior, and informed Indigenous consent. Require that representatives on a Cultural IRB represent a diversity of viewpoints, and prioritize the appointment of cultural knowledge holders (*e.g.*, kūpuna in a Hawaiian context). Ensure the factors assessed are relevant to the communities affected by the project; *e.g.*, similar initiatives in Hawai'i include assessments of family and community life, human well-being and spirituality, natural environment and cultural and ecological resources, customs and practices, Indigenous and common law rights, and the economic well-being of Hawaiians[17].

### 3.3. Require that every observational astronomer who uses Maunakea learn Hawaiian history and culture, regardless of whether they are physically present in Hawai'i.

**Problem 3:** Many astronomers residing in Hawai'i have cursory knowledge of Hawaiian history, culture and the context of Maunakea conflicts. For example, much of the IfA website[x] is written from a non-Indigenous perspective, and discusses only the scientific context for Maunakea development. Failing to mention Native Hawaiian opposition reinforces a widespread phenomenon of Indigenous erasure, which is further is exacerbated by the remote nature of modern astronomy: the majority of telescopes on Maunakea use remote observing, such that many astronomers access observations without ever visiting the Mauna. While efficient, this physical disconnection between researchers and communities alienates astronomers and Native Hawaiians, obscures ongoing settler colonialism in Hawai'i, and perpetuates a lack of awareness of concerns and views of many Native Hawaiians.

**Recommendation 3:** We recommend that the American Astronomical Society (AAS) and international partners work in collaboration with the Maunakea telescopes and Hawaiian kūpuna and educators to develop courses or workshops on Hawaiian and Indigenous history and knowledge systems for both Hawai'i-based astronomers and those accessing observations remotely. The curriculum should be required of all Hawai'i-based astronomers at all career stages as well as *every* astronomer who obtains observations from a telescope on Maunakea. Similar to data acquisition, course context could be regularly delivered via online and/or in-person workshops to university partners. Finally, these course materials should be accessible at any time, and additional classes on Hawaiian history, culture, and literature should be made available during the yearly AAS meetings. As long as there is a benefit from telescopes, **each university has a substantial responsibility to invest in perpetuating education on Hawaiian history and culture and engage in practices of reciprocity with living Hawaiian communities.**

The development of these materials must be a Native Hawaiian-led initiative and in alignment with established methods of knowledge co-production[18]. Ideally, the materials and workshop facilitators should enable open discussion of conflicts and the complex web of relationships between Native Hawaiians, University of Hawai'i, IfA, the Maunakea Observatories, and the state of Hawai'i to place modern disputes into context. Engaging in open-ended discussions about relationship norms between individuals and institutions alike will allow for better relational accountability among different groups and contribute to equitable astronomy research and education. Relationship documents could be created to make explicit the roles, responsibilities, contributions, and dissolvement procedures for astronomy projects on Indigenous land.

Finally, we recommend that Maunakea outreach and informational materials be explicit about the historical context of professional astronomy's presence in the Hawaiian Islands and beyond. We suggest that the AAS and all Maunakea observatories adopt a land acknowledgement as a first step towards acknowledging professional astronomy's history in the Hawaiian Islands[xi].

### 3.4. Establish equitable, iterative dialogue with Native Hawaiians.

**Problem 4:** Current outreach efforts are not sufficient as long as there continues to be a lack of dialogue between astronomers and Native Hawaiians *as equals*. To date, astronomers have centered a small portion of Native Hawaiian viewpoints, which has damaged relationships both between and within these communities.

---

[x] IfA website page "History of astronomy in Hawai'i": http://www.ifa.hawaii.edu/ifa/history.shtml
[xi] Example guidelines on land acknowledgements: https://native-land.ca/territory-acknowledgement/

**Recommendation 4:** Establish healthy, collaborative dialogue with the Hawaiian community, and seek out opportunities to rebuild trust. Build Native Hawaiian feedback directly into astronomy projects. For example, telescope projects could support a paid consultant position for cultural knowledge holders similar to the Kupuna program in Hawaiʻi elementary schools[19]. While *this would not supersede community feedback*, it could serve as an additional point of connection between astronomers and Native Hawaiians.

### 3.5. Build support for Native Hawaiian astronomy students.

**Problem 5:** As in other STEM fields, Indigenous peoples remain severely underrepresented in astronomy. To date, only three Native Hawaiians have earned a PhD in Physics, and only one holds a PhD in Astronomy. Collectively, there are less than 20 Indigenous physicists and astronomers with PhDs worldwide, and **there are no Indigenous tenure-track faculty at leading research institutions—*i.e.*, the primary stakeholders in developing astronomical instrumentation.** Poor Indigenous representation is exacerbated by limited support structures for post-graduate trainees and early career Native Hawaiian scientists; funneling more marginalized people into the existing career pathways will not fix its current "leakiness"[20]. Importantly, the lack of institutional support for meaningful, consent-driven, and reciprocal engagement with Indigenous peoples may deepen the perception that outreach efforts are performed in bad faith. Current astronomy outreach efforts, while important, are an insufficient remedy to the astronomy community's now strained relationship with Native Hawaiians.

**Recommendation 5:** Financially invest in a welcoming and supportive environment for Native Hawaiian and Indigenous astronomers by incentivizing culturally-aware student training and mentorship. This can be done in a number of ways; *e.g.*, it could begin by identifying and coalescing an international working group or network of Indigenous astronomers and educators familiar with both the specifics of their disciplines and best practices in Indigenous pedagogies and student mentorship. We suggest that such a network could exchange ideas that speak to specific complexities and challenges within their local communities, but could also work to identify more general cross-cutting themes and practices.

### 3.6. Develop astronomy-specific ethical guidelines and accountability structures

**Problem 6:** Astronomers utilizing observatories on Maunakea benefit directly from control of the summit by the State and the University of Hawaiʻi. While many astronomers are likely unaware of the specifics of colonialism in Hawaiʻi, the ongoing harm to Native Hawaiians—and, by extension, to Indigenous peoples—should be addressed as it remains a source of unacknowledged inequity and disparity.

**Recommendation 6:** One path towards restorative justice may involve working towards reciprocal relationships among professional astronomers, institutions with jurisdiction over the Maunakea summit, and the Native Hawaiian community at large. **As a first step, the astronomy community must develop a code of conduct that addresses colonization as part of its ethical considerations, and should work with Indigenous communities to co-generate context-specific "best practice" guidelines.** They should focus not only on research and outreach, but also on research infrastructure, land tenure, and access--the key places where astronomers interface with Indigenous communities. A number of already-established best practices from across other disciplines that may serve as a useful springboard[16,21–23].

### 3.7. Funding agencies must hold PIs accountable for the research environments they create and perpetuate.

**Problem 7:** There is currently no evaluation process conducted by funding agencies to ensure that grant awardees are held accountable for the research environments they create. Few institutional structures exist to protect trainees and junior scientists from toxic mentors and collaborators. Destructive behavior, especially towards Indigenous and minority scholars (*e.g.,* section 2.2 above), easily goes unchecked. Faculty have little structural incentive to evaluate their behavior or develop greater professionalism. The current funding model relies heavily on transient initiatives set up by students, post-doctoral scholars, or temporary employees; most are opt-in traineeships with few faculty participants.

**Recommendation 7: We recommend that funding agencies hold grant awardees accountable to ethical guidelines** by developing initiatives to evaluate the health or hostility that exists within each research group. Funding agencies should reward Principal Investigators (PIs) who actively promote healthy and inclusive research environments, and should not reward PIs who perpetuate hostility, regardless of their research output. **Funding agencies' assessments of "quality of research" should include markers of research conduct.** Without deliberate steps towards inclusionary practices, truly diverse research environments will remain out of reach.

## 4. Long-term recommendations: Diversify the culture of science

Below provide a number of broader suggestions for astronomers aiming to improve diversity and inclusion initiatives aimed at non-astronomers and Indigenous peoples. **We suggest that these recommendations be incorporated into scientific training and project development beyond the end of the current decadal assessment cycle.** These recommendations are expanded from an essay[24] published by an astronomer after the start of the 2019 TMT demonstrations.

**1. Establish ongoing and reciprocal dialogue between scientists and Indigenous cultural knowledge holders.** Build relationships centered on openness and trust. When developing projects on Indigenous land, work to gain informed and ongoing consent, and respect the right of Indigenous people to reject projects at any stage.

**2. Learn the history of colonization in science to gain context for modern interactions.** This includes a critical examination of the practice and process of science—e.g., how has astronomy been done in the past, by both Indigenous and colonizing groups, and what are the modern social impacts of the legacy of Western science? Recognize that scientific institutions have benefitted from the disenfranchisement of Indigenous and minority groups, and that the legacy of colonialism through the practice of science remains a strong barrier to Indigenous and minority engagement with the scientific community[16]. To begin, read and listen to Native Hawaiian critiques of the TMT project and of astronomy on Maunakea. Remember that many critiques will not be published through traditional scientific channels, but will rather occur through news media (opinion pieces, editorials, and collective writing) and in non-academic outlets (e.g., community board meetings, verbal discourse, videos, social media).

**3. Learn and connect with Indigenous knowledge systems.** Indigenous nations and peoples have explored the world and Universe for far longer than astronomers have had telescopes. These knowledges are not uniform, nor is there only one Indigenous knowledge. However, Indigenous knowledges tend to build on a deep connection with the land, the water and the sky, and contain axioms that vary from mainstream scientific perspectives. In Hawaiʻi, astronomers can learn the Native Hawaiian values of aloha ʻāina, malama ʻāina, and kuleana to build an understanding of the significance of Mauna a Wākea.

**4. Consider what we and others value.** A key component to relationship-building is coming to an understanding that different groups employ different value systems, and that closely-held values will steer the decision-making process. Science itself is an ideology—a value system—that can inform decisions, just as Indigenous cultures and knowledges carry and transmit specific values.

**5. Emphasize relational skills in scientific training.** Relational communication takes many forms, but in science circles can focus on surfacing mutual values and needs in communication, project design, and implementation[25]. At present, scientific training does not emphasize scientists' abilities to work with conflict, engage with emotional responses, and take into account the value systems held by individuals and communities outside of the scientific project team's inner circles. Yet most scientists, including astronomers, have long engaged with complex systems, and these skills can be targeted towards navigating ambiguous or emotional topics. Training in relational practice can include an emphasis on deep listening, valuing and sharing personal experiences, employing introspection and self-reflection, and accountability through iterative communication with non-astronomer groups.

## 5. Conclusion

The situation on Maunakea provides an opportunity to examine our relationships, especially among the intersecting groups of Indigenous people and the scientific community. In this paper, we have outlined how the potential construction of the Thirty Meter Telescope is a point of extreme tension. Protectors at the base of Maunakea remain steadfast in their commitment to prioritize the well-being of land above their own physical safety—so much so that some are prepared to die in order to stop TMT construction. Demonstrations in solidarity are continuously being held across the state of Hawai'i and around the world. Given that there is no Indigenous consent for TMT, **the project must halt construction, reconsider its position, and restart the process of engaging in reciprocal dialogue with Native Hawaiians.** Crucially, the TMT project must enter into these negotiations willing to accept the choice made by Hawai'i's Indigenous people, even if this means that the project must withdraw and consider an alternate location.

The astronomical community must take this conversation very seriously. At this moment, we have an opportunity to shape the future of the field, and to work towards a practice of science that is truly ethical—one that upholds human and Indigenous rights. We are at a point in history where the construction of large-scale scientific instruments requires re-evaluation of the way in which the field of astronomy engages with local and Indigenous communities. Our present actions will inform the processes through which we construct future instrumentation—thus, we must carefully consider the values we hope to promote. The recommendations we outline can serve as first steps towards building reciprocal and *equal* relationships between astronomers and the Indigenous people on whose land they work. Ethical science is predicated on and informed by the values and morals of society, including those that may be beyond Western traditions[26]. In upholding the core Hawaiian values of kapu aloha and aloha 'āina in our practice of science, we can reaffirm our commitment to an ethical scientific practice.

We thank Ellis Avallone for constructive feedback and support. A hiki i ke aloha ʻāina hope loa.

**References**

1. Meyer, M. A. *Maintaining a kapu aloha for Mauna Kea: understanding Mauna, culture, and intention through moana-nui-ākea.* (2015).
2. Noe Tanigawa. Kapu aloha: Power of love. *HPR* (2015).
3. Casumbal-Salazar, I. A Fictive Kinship: Making "Modernity," "Ancient Hawaiians," and the Telescopes on Mauna Kea. *Nativ. Am. Indig. Stud.* **4**, 1–31 (2017).
4. Goodyear-Kaʻōpua, N. Protectors of the Future, Not Protestors of the Past: Indigenous Pacific Activism and Mauna a Wākea. *South Atl. Q.* **116**, 184–194 (2017).
5. *Understanding Maunakea: A primer on cultural and environmental impacts.* (2019).
6. Nordstrom, G. Review of Mauna Kea: Temple Under Siege (documentary film). (2006).
7. Auditor, S. of H. Audit of the Management of Mauna Kea and the Mauna Kea Science Reserve and the Legislature of the State of Hawaiʻi. (1998).
8. Stemwedel, J. D. The Thirty Meter Telescope Reveals Ethical Challenges For The Astronomy Community. *Forbes* (2015).
9. Staff. UH denounces professor's 'hurtful' statements about Kamehameha Schools students. *Hawaiʻi News Now* (2019).
10. Simons, D. Astro 2020 APC White Paper: The Future of Maunakea Astronomy. (2019).
11. Kimura, K. *et al.* Astro 2020 APC White Paper: A Hua He Inoa: Hawaiian Culture-Based Celestial Naming. (2019).
12. Begay, D. *et al.* Astro2020 APC White Paper Collaboration with Integrity: Indigenous Knowledge in 21st Century Astronomy. (2019).
13. Hall, S. Why care about astronomy? *Universe Today* (2014).
14. Board, S. S. & Council, N. R. *Portals to the Universe: The NASA Astronomy Science Centers.* (National Academies Press, 2007).
15. Cominsky, L. R. Education and public outreach in astronomy and beyond. *Nat. Astron.* **2**, 14–15 (2018).
16. Smith, L. T. *Decolonizing Methodologies.* (London ; New York : Zed Books ; Dunedin : University of Otago Press ; New York : distributed in the USA exclusively by St Martin's Press, 1999., 1999).
17. Matsuoka, J., McGregor, D. & Minberi, L. Native Hawaiian Cultural Impact Assessment Workbook. in *Hawaii Externalities Workbook* (1997).
18. Wyborn, C. *et al.* Co-Producing Sustainability: Reordering the Governance of Science, Policy, and Practice. *Annu. Rev. Environ. Resour.* **44**, 319–346 (2019).
19. Kaomea, J. Dilemmas of an Indigenous Academic: A Native Hawaiian Story. *Contemp. Issues Early Child.* **2**, 67–82 (2001).
20. Flaherty, K. The Leaky Pipeline for Postdocs: A study of the time between receiving a PhD and securing a faculty job for male and female astronomers. (2018).
21. David-Chavez, D. M. & Gavin, M. C. A global assessment of Indigenous community engagement in climate research. *Environ. Res. Lett.* **13**, 123005–123018 (2018).
22. Morishige, K. *et al.* Nā Kilo ʻĀina: Visions of Biocultural Restoration through Indigenous Relationships between People and Place. *Sustainability* **10**, 3320–3368 (2018).
23. Mistry, J. & Berardi, A. Bridging indigenous and scientific knowledge. *Science (80-. ).* **352**, 20150173–20150174 (2016).
24. de los Reyes, M. Where do we go from here? Small ways to make the field of astronomy a better place. *Medium* (2019).
25. Kearns, F. R. A Relational Approach to Climate Change BT  - Climate Change Across the Curriculum. in *Climate Change Across the Curriculum* (ed. Fretz, E.) 219–234 (Lexington Books, 2015).
26. Alegado, R. Telescope opponents fight the process, not science. *Nature* **572**, 7 (2019).